\documentclass[prl,onecolumn,nofootinbib]{revtex4}
\usepackage{bm,amsmath,amssymb,graphicx}

\begin{document}

\title{Intrinsic spin requires gravity with torsion and curvature}
\author{Nikodem Pop{\l}awski}
\affiliation{Department of Physics, Indiana University, Bloomington, Indiana, USA}

\begin{abstract}
We show that the intrinsic angular momentum of matter in curved spacetime requires the metric-affine formulation of gravity, in which the antisymmetric part of the affine connection (the torsion tensor) is not constrained to be zero but is a variable in the principle of stationary action.
Regarding the tetrad and spin connection (or the metric and torsion tensors) as independent variables gives the correct generalization of the conservation law for the total (orbital plus intrinsic) angular momentum to the presence of the gravitational field.
The metric-affine formulation extends general relativity to the simplest theory of gravity with intrinsic spin: the Einstein-Cartan-Sciama-Kibble theory.
We also show that teleparallel gravity, which constrains the connection by setting the curvature tensor to zero, is inconsistent with the conservation of the total angular momentum.
\end{abstract}

\maketitle

\paragraph{Angular momentum without gravitational field.}
We consider a physical system in the absence of the gravitational field, described by a matter Lagrangian density $\mathfrak{L}$ which depends on matter fields $\phi$, their first partial derivatives $\phi_{,i}$ with respect to the coordinates $x^i$, and $x^i$ \cite{Niko,KS}.
Under an infinitesimal coordinate transformation $x^i\rightarrow x^{'i}=x^i+\xi^i$, where $\xi^i=\delta x^i$ is a variation of $x^i$, the Lagrangian density $\mathfrak{L}$ transforms like a scalar density: $\delta \mathfrak{L}=(|\partial x^i/\partial x^{'i}|-1)\mathfrak{L}=-\xi^i_{\phantom{i},i}\mathfrak{L}$.
The variation $\delta\mathfrak{L}$ is also equal to $\delta\mathfrak{L}=(\partial\mathfrak{L}/\partial\phi)\delta\phi+\bigl(\partial\mathfrak{L}/\partial(\phi_{,i})\bigr)\delta(\phi_{,i})+(\bar{\partial}\mathfrak{L}/\partial x^i)\xi^i$, where $\bar{\partial}$ denotes partial differentiation with respect to $x^i$ at constant $\phi$ and $\phi_{,i}$.
Using the Lagrange equations $\partial\mathfrak{L}/\partial\phi-\partial_i\bigl(\partial\mathfrak{L}/\partial(\phi_{,i})\bigr)=0$, and the identities $\mathfrak{L}_{,i}=\bar{\partial}\mathfrak{L}/\partial x^i+(\partial\mathfrak{L}/\partial\phi)\phi_{,i}+\bigl(\partial\mathfrak{L}/\partial(\phi_{,j})\bigr)\phi_{,ji}$ and $\delta(\phi_{,i})=(\delta\phi)_{,i}-\xi^j_{\phantom{j},i}\phi_{,j}$, leads to the conservation law:
\begin{equation}
\mathfrak{J}^i_{\phantom{i},i}=0,
\label{Noe1}
\end{equation}
for the current vector density
\begin{equation}
\mathfrak{J}^i=\xi^i\mathfrak{L}+\frac{\partial\mathfrak{L}}{\partial(\phi_{,i})}(\delta\phi-\xi^j\phi_{,j}).
\label{Noe2}
\end{equation}
The existence of a conservation law for each continuous symmetry of a Lagrangian density formulates the Noether theorem.

For Lorentz rotations, we have $\xi^i=\epsilon^i_{\phantom{i}j}x^j$ and $\delta\phi=(1/2)\epsilon_{ij}G^{ij}\phi$, where $\epsilon_{ij}=-\epsilon_{ji}$ are infinitesimal quantities and $G^{ij}$ are the generators of the Lorentz group.
The corresponding current (\ref{Noe2}) is
\begin{equation}
\mathfrak{J}^i=\epsilon^{kl}\biggl(\frac{\partial\mathfrak{L}}{\partial(\phi_{,i})}\phi_{,[l}x_{k]}-\delta^i_{[l}x_{k]}\mathfrak{L}+\frac{1}{2}\frac{\partial\mathfrak{L}}{\partial(\phi_{,i})}G_{kl}\phi\biggr),
\end{equation}
where $[\,]$ denotes antisymmetrization.
Because $\epsilon^{kl}$ are arbitrary, (\ref{Noe1}) gives the conservation law:
\begin{equation}
\mathfrak{M}_{kl\phantom{i},i}^{\phantom{kl}i}=0,
\label{amd1}
\end{equation}
for the angular momentum density
\begin{equation}
\mathfrak{M}_{kl}^{\phantom{kl}i}=x_k\theta^{\phantom{l}i}_l-x_l\theta^{\phantom{k}i}_k+\frac{\partial\mathfrak{L}}{\partial(\phi_{,i})}G_{kl}\phi,
\label{amd2}
\end{equation}
where 
\begin{equation}
\theta^{\phantom{i}k}_i=\frac{\partial\mathfrak{L}}{\partial(\phi_{,k})}\phi_{,i}-\delta^k_i \mathfrak{L}
\label{cemd1}
\end{equation}
is the canonical energy-momentum density.
The first two terms on the right-hand side of (\ref{amd2}) form the orbital angular momentum density, and the last term is the canonical spin density $\Sigma_{kl}^{\phantom{kl}i}$.

For translations, where $x^i$ are Cartesian coordinates, we have $\xi^i=\epsilon^i=$ const and $\delta\phi=0$.
The current (\ref{Noe2}) is $\mathfrak{J}^i=\epsilon^i\mathfrak{L}-(\partial\mathfrak{L}/\partial\phi_{,i})\epsilon^j\phi_{,j}$, so the conservation law (\ref{Noe1}) gives $\epsilon^j\theta^{\phantom{j}i}_{j\phantom{i},i}=0$.
Because $\epsilon^i$ are arbitrary, this relation gives the conservation law for the canonical energy-momentum density (\ref{cemd1}):
\begin{equation}
\theta^{\phantom{i}j}_{i\phantom{j},j}=0.
\label{cemd2}
\end{equation}
This law also results from differentiating $\mathfrak{L}$ over $x^i$ and using the Lagrange equations.
The conservation law (\ref{amd1}) for the total angular momentum density can be written, using (\ref{cemd2}), as the conservation law for the canonical spin density in the special theory of relativity \cite{Niko,KS}:
\begin{equation}
\Sigma_{kl\phantom{i},i}^{\phantom{kl}i}=\theta_{kl}-\theta_{lk}.
\label{amd3}
\end{equation}

\paragraph{Angular momentum with gravitational field.}
In the metric-affine formulation of gravity, the tetrad $e^i_a$ and the spin connection
\begin{equation}
\omega^a_{\phantom{a}bk}=e^a_j e^j_{\phantom{j}b;k}=e^a_j(e^j_{\phantom{j}b,k}+\Gamma^{\,\,j}_{i\,k}e^i_b)
\label{spincon}
\end{equation}
are dynamical variables describing the geometry of spacetime \cite{Niko,KS,Hehl,rev}.
Semicolon denotes the covariant derivative with respect to the affine connection $\Gamma^{\,\,i}_{j\,k}$.
The affine connection is asymmetric in the lower indices and its antisymmetric part is the torsion tensor \cite{Niko,KS,Hehl,rev}:
\begin{equation}
S^i_{\phantom{i}jk}=\Gamma^{\,\,\,\,i}_{[j\,k]}.
\end{equation}
The spin connection appears in the covariant derivative of a Lorentz vector: $V^a_{\phantom{a}|i}=V^a_{\phantom{a},i}+\omega^a_{\phantom{a}bi}V^b$ and $V_{a|i}=V_{a,i}-\omega^b_{\phantom{b}ai}V_b$, analogously to $\Gamma^{\,\,i}_{j\,k}$ in the covariant derivative of a vector, $V^k_{\phantom{k};i}=V^k_{\phantom{k},i}+\Gamma^{\,\,k}_{l\,i}V^l$ and $V_{k;i}=V_{k,i}-\Gamma^{\,\,l}_{k\,i}V_l$.
The tetrad relates spacetime coordinates $i,j,...$ to local Lorentz coordinates $a,b,...$: $V^a=V^i e^a_i$.
Its covariant derivative vanishes by means of (\ref{spincon}): $e^a_{i|k}=e^a_{i,k}-\Gamma^{\,\,j}_{i\,k}e^a_j+\omega^a_{\phantom{a}bk}e^b_i=0$, where vertical bar denotes the covariant derivative acting on both spacetime and Lorentz coordinates.
Lorentz coordinates are thus lowered or raised by the Minkowski metric tensor $\eta_{ab}$ of a flat spacetime, analogously to the metric tensor $g_{ik}$ lowering or raising spacetime coordinates.
The metricity condition $g_{ij;k}=0$ gives the affine connection $\Gamma^{\,\,k}_{i\,j}=\{^{\,\,k}_{i\,j}\}+C^k_{\phantom{k}ij}$, where $\{^{\,\,k}_{i\,j}\}=(1/2)g^{km}(g_{mi,j}+g_{mj,i}-g_{ij,m})$ are the Christoffel symbols, $C^i_{\phantom{i}jk}=S^i_{\phantom{i}jk}+2S_{(jk)}^{\phantom{(jk)}i}$ is the contortion tensor, and $(\,)$ denotes symmetrization.
It also constrains the spin connection to be antisymmetric in its Lorentz indices: $\omega^{ab}_{\phantom{ab}i}=-\omega^{ba}_{\phantom{ba}i}$.
Instead of $e^i_a$ and $\omega^{ab}_{\phantom{ab}i}$, the metric tensor $g_{ik}=\eta_{ab}e^a_i e^b_k$ and the torsion tensor $S^j_{\phantom{j}ik}=\omega^j_{\phantom{j}[ik]}+e^a_{[i,k]}e^j_a$ can be taken as the dynamical variables.

We consider a physical system in the presence of the gravitational field.
The variation of $\mathfrak{L}_\textrm{m}$ with respect to the spin connection defines the dynamical spin density \cite{Niko,KS,Hehl,rev}
\begin{equation}
\mathfrak{S}_{ab}^{\phantom{ab}i}=2\frac{\delta\mathfrak{L}_\textrm{m}}{\delta\omega^{ab}_{\phantom{ab}i}}=2\frac{\partial\mathfrak{L}_\textrm{m}}{\partial\omega^{ab}_{\phantom{ab}i}},
\label{spind1}
\end{equation}
which is antisymmetric in the Lorentz indices: $\mathfrak{S}_{ab}^{\phantom{ab}i}=-\mathfrak{S}_{ba}^{\phantom{ba}i}$.
The spin tensor is defined as $s_{ijk}=\frac{1}{\mathfrak{e}}\mathfrak{S}_{ijk}$, where $\mathfrak{e}=\mbox{det}\,e^a_i=\sqrt{-\mbox{det}\,g_{ik}}$.
The second equality in (\ref{spind1}) is satisfied because a matter Lagrangian density may depend on the spin connection but not on its derivatives; a scalar density depending on derivatives of $\omega^{ab}_{\phantom{ab}i}$ is a Lagrangian density for the gravitational field.
The spin density is also given by
\begin{equation}
\mathfrak{S}_{ij}^{\phantom{ij}k}=2\frac{\delta\mathfrak{L}_\textrm{m}}{\delta C^{ij}_{\phantom{ij}k}}=2\frac{\partial\mathfrak{L}_\textrm{m}}{\partial C^{ij}_{\phantom{ij}k}}.
\label{spind2}
\end{equation}
The variation of the Lagrangian density for matter $\mathfrak{L}_\textrm{m}$ with respect to the tetrad defines the dynamical energy-momentum density \cite{Niko,KS,Hehl,rev}
\begin{equation}
\mathfrak{T}^{\phantom{i}a}_i=\frac{\delta\mathfrak{L}_\textrm{m}}{\delta e^i_a}=\frac{\partial\mathfrak{L}_\textrm{m}}{\partial e^i_a}-\partial_j\biggl(\frac{\partial\mathfrak{L}_\textrm{m}}{\partial(e^i_{a,j})}\biggr).
\label{demd1}
\end{equation}
The metric energy-momentum tensor $T_{ij}=(2/\mathfrak{e})(\delta\mathfrak{L}_\textrm{m}/\delta g^{ij})=(2/\mathfrak{e})[\partial\mathfrak{L}_\textrm{m}/\partial g^{ij}-\partial_k\bigl(\partial\mathfrak{L}_\textrm{m}/\partial(g^{ij}_{\phantom{ij},k})\bigr)]$ is symmetric, $T_{ij}=T_{ji}$.
It is related to the dynamical energy-momentum density and the spin tensor by the Belinfante-Rosenfeld relation: $T_{ik}=\mathfrak{T}_{ik}/\mathfrak{e}-(1/2)(s_{ik}^{\phantom{ik}j}-s_{k\phantom{j}i}^{\phantom{k}j}+s^j_{\phantom{j}ik})_{;j}+S_j(s_{ik}^{\phantom{ik}j}-s_{k\phantom{j}i}^{\phantom{k}j}+s^j_{\phantom{j}ik})$, where $S_i=S^k_{\phantom{k}ik}$ is the torsion vector.
Since the variations $\delta\omega^{ab}_{\phantom{ab}i}$ are independent of $\delta e^i_a$, the spin density is independent of the energy-momentum density.

The Lorentz group is the group of tetrad rotations, $e^a_i=\Lambda^a_{\phantom{a}b}e^b_i$, where $\Lambda^a_{\phantom{a}b}$ is a Lorentz matrix.
Since a matter Lagrangian density $\mathfrak{L}_\textrm{m}(\phi,\phi_{,i})$ is invariant under local, proper Lorentz transformations, it is invariant under tetrad rotations: $\delta\mathfrak{L}_\textrm{m}=(\partial\mathfrak{L}_\textrm{m}/\partial\phi)\delta\phi+\bigl(\partial\mathfrak{L}_\textrm{m}/\partial(\phi_{,i})\bigr)\delta(\phi_{,i})+\mathfrak{T}^{\phantom{i}a}_i\delta e^i_a+(1/2)\mathfrak{S}_{ab}^{\phantom{ab}i}\delta\omega^{ab}_{\phantom{ab}i}=0$, where the changes $\delta$ are caused by a tetrad rotation.
Upon integration of $\delta\mathfrak{L}_\textrm{m}$ over spacetime, the terms with $\phi$ and $\phi_{,i}$ vanish because of the Lagrange equations:
\begin{equation}
\int\Bigl(\mathfrak{T}^{\phantom{i}a}_i\delta e^i_a+\frac{1}{2}\mathfrak{S}_{ab}^{\phantom{ab}i}\delta\omega^{ab}_{\phantom{ab}i}\Bigr)d\Omega=0.
\label{var}
\end{equation}
For an infinitesimal Lorentz transformation, $\Lambda^a_{\phantom{a}b}=\delta^a_b+\epsilon^a_{\phantom{a}b}$, where $\epsilon^a_{\phantom{a}b}=-\epsilon_b^{\phantom{b}a}$ are infinitesimal quantities, the tetrad $e^a_i$ changes by $\delta e^a_i=\Lambda^a_{\phantom{a}b}e^b_i-e^a_i=\epsilon^a_{\phantom{a}i}$, and the tetrad $e_a^i$ changes by $\delta e_a^i=-\epsilon_{\phantom{i}a}^i$ because of $e^a_i e_a^j=\delta_i^j$.
Accordingly, the spin connection changes by $\delta\omega^{ab}_{\phantom{ab}i}=\delta(e^a_j \omega^{j b}_{\phantom{j b}i})=\epsilon^a_{\phantom{a}j}\omega^{j b}_{\phantom{j b}i}-e^a_j \epsilon^{j b}_{\phantom{j b};i}=\epsilon^a_{\phantom{a}c}\omega^{cb}_{\phantom{cb}i}-e^a_j \epsilon^{j b}_{\phantom{j b}|i}+\epsilon^a_{\phantom{a}c}\omega^{bc}_{\phantom{bc}i}=-\epsilon^{ab}_{\phantom{ab}|i}$.
Substituting these variations into (\ref{var}) and using partial integration $\int\mathfrak{V}^i_{\phantom{i};i}d\Omega=2\int S_i \mathfrak{V}^i d\Omega$, where $\mathfrak{V}$ is any contravariant vector density, leads to $-\int\Bigl(\mathfrak{T}^{\phantom{i}a}_i\epsilon^i_{\phantom{i}a}+\frac{1}{2}\mathfrak{S}_{ab}^{\phantom{ab}i}\epsilon^{ab}_{\phantom{ab}|i}\Bigr)d\Omega=-\int\Bigl(\mathfrak{T}_{ij}\epsilon^{ij}+\frac{1}{2}\mathfrak{S}_{ij}^{\phantom{ij}k}\epsilon^{ij}_{\phantom{ij}|k}\Bigr)d\Omega=\int\Bigl(-\mathfrak{T}_{[ij]}-S_k\mathfrak{S}_{ij}^{\phantom{ij}k}+\frac{1}{2}\mathfrak{S}_{ij\phantom{k};k}^{\phantom{ij}k}\Bigr)\epsilon^{ij}d\Omega=0$.
Since the infinitesimal Lorentz rotation $\epsilon^{ij}$ is arbitrary, we obtain the conservation law for the spin density \cite{Niko,KS,Hehl,rev}:
\begin{equation}
\mathfrak{S}_{ij\phantom{k};k}^{\phantom{ij}k}-2S_k\mathfrak{S}_{ij}^{\phantom{ij}k}=\mathfrak{T}_{ij}-\mathfrak{T}_{ji}.
\label{amd4}
\end{equation}
This law can be written as $\mathfrak{S}^{ijk}_{\phantom{ijk},k}-\Gamma^{\,\,i}_{l\,k}\mathfrak{S}^{jlk}+\Gamma^{\,\,j}_{l\,k}\mathfrak{S}^{ilk}-2\mathfrak{T}^{[ij]}=0$.
The conservation law (\ref{amd4}) also results from antisymmetrizing the Belinfante-Rosenfeld relation with respect to the indices $i,k$.

A matter Lagrangian density $\mathfrak{L}_\textrm{m}$ can be written as $\mathfrak{L}_\textrm{m}=\mathfrak{e}L$, where $L$ is a scalar.
If $\mathfrak{L}_\textrm{m}$ depends on matter fields $\phi$ (minimally coupled to the affine connection) and their first derivatives $\phi_{,i}$, and the fields $\phi$ do not contain vector indices, then the tetrad appears in $L$ only through derivatives of $\phi$, in a covariant combination $e^i_a \phi_{|i}$.
Such fields can be, for example, spinor fields.
Varying $\mathfrak{L}_\textrm{m}$ with respect to the tetrad gives $\delta\mathfrak{L}_\textrm{m}=\mathfrak{e}\delta L-\mathfrak{e}e^a_i L\delta e^i_a=\mathfrak{e}\bigl(\partial L/\partial(\phi_{|a})\bigr)\phi_{|i}\delta e^i_a-\mathfrak{L}_\textrm{m}e^a_i \delta e^i_a=\Bigl(\bigl(\partial\mathfrak{L}_\textrm{m}/\partial(\phi_{|a})\bigr)\phi_{|i}-e^a_i\mathfrak{L}_\textrm{m}\Bigr)\delta e^i_a$.
The dynamical energy-momentum density (\ref{demd1}) is therefore $\mathfrak{T}^{\phantom{i}a}_i=\bigl(\partial\mathfrak{L}_\textrm{m}/\partial(\phi_{|a})\bigr)\phi_{|i}-e^a_i \mathfrak{L}_\textrm{m}$.
The corresponding tensor with two coordinate indices,
\begin{equation}
\mathfrak{T}^{\phantom{i}k}_i=\frac{\partial\mathfrak{L}_\textrm{m}}{\partial(\phi_{|k})}\phi_{|i}-\delta^k_i \mathfrak{L}_\textrm{m}=\frac{\partial\mathfrak{L}_\textrm{m}}{\partial(\phi_{,k})}\phi_{|i}-\delta^k_i \mathfrak{L}_\textrm{m},
\label{cemd3}
\end{equation}
generalizes the canonical energy-momentum density (\ref{cemd1}) to the presence of the gravitational field \cite{Niko,rev}.
The spin connection $\omega^{ab}_{\phantom{ab}i}$ appears in $\mathfrak{L}_\textrm{m}$ only through derivatives of $\phi$, in a combination $-\bigl(\partial\mathfrak{L}_\textrm{m}/\partial(\phi_{,i})\bigr)\Gamma_i\phi$, where $\Gamma_i=-(1/2)\omega_{abi}G^{ab}$ is the connection in the covariant derivative of $\phi$: $\phi_{|i}=\phi_{,i}-\Gamma_i\phi$.
The dynamical spin density (\ref{spind1}) is therefore $\mathfrak{S}_{ab}^{\phantom{ab}i}=\bigl(\partial\mathfrak{L}_\textrm{m}/\partial(\phi_{,i})\bigr)G_{ab}\phi$.
The corresponding tensor with two coordinate indices, $\mathfrak{S}_{kl}^{\phantom{kl}i}=\bigl(\partial\mathfrak{L}_\textrm{m}/\partial(\phi_{,i})\bigr)G_{kl}\phi$, coincides with the canonical spin density $\Sigma_{kl}^{\phantom{kl}i}$ in (\ref{amd2}).
Consequently, the conservation law (\ref{amd4}) for the spin density generalizes (\ref{amd3}) to the presence of the gravitational field \cite{Niko,KS}.

In the metric formulation of gravity, the tetrad (or the metric tensor) is the only dynamical variable representing the gravitational field \cite{LL}.
In that formulation, the torsion tensor is constrained to be zero, so the affine connection is equal to the Levi-Civita connection given by the Christoffel symbols: $\Gamma^{\,\,k}_{i\,j}=\{^{\,\,k}_{i\,j}\}$.
Accordingly, the spin connection is a function of the tetrad and its first derivatives, so the variations $\delta\omega^{ab}_{\phantom{ab}i}$ are functions of $\delta e^i_a$ and their derivatives.
The spin density (\ref{spind1}) is thus a function of the energy-momentum density, forming a part of the orbital angular momentum density, whereas (\ref{spind2}) is no longer valid.
The relation (\ref{var}) reduces to $\int\mathfrak{T}^{\phantom{i}a}_i\delta e^i_a d\Omega=0$ and (\ref{amd4}) reduces to $\mathfrak{T}_{ij}=\mathfrak{T}_{ji}$, which is not a generalization of (\ref{amd3}) unless the intrinsic spin vanishes.
The metric formulation therefore excludes the intrinsic spin.
Consequently, the observed existence of matter with intrinsic spin requires the metric-affine formulation and a nonzero torsion tensor.
For example, metric and torsionless $f(R)$ gravity theories \cite{fR} are ruled out.

\paragraph{Metric-affine gravity: Einstein-Cartan-Sciama-Kibble theory.}
The Lagrangian density for the gravitational field contains the first derivatives of the spin or affine connection, which appear through the curvature tensor, $R^a_{\phantom{a}bij}=\omega^a_{\phantom{a}bj,i}-\omega^a_{\phantom{a}bi,j}+\omega^a_{\phantom{a}ci}\omega^c_{\phantom{c}bj}-\omega^a_{\phantom{a}cj}\omega^c_{\phantom{c}bi}$ or $R^i_{\phantom{i}mjk}=\partial_{j}\Gamma^{\,\,i}_{m\,k}-\partial_{k}\Gamma^{\,\,i}_{m\,j}+\Gamma^{\,\,i}_{l\,j}\Gamma^{\,\,l}_{m\,k}-\Gamma^{\,\,i}_{l\,k}\Gamma^{\,\,l}_{m\,j}$ \cite{Niko,KS,rev}.
This tensor satisfies the Bianchi identity, $R^i_{\phantom{i}n[jk;l]}=2R^i_{\phantom{i}nm[j}S^m_{\phantom{m}kl]}$, and the cyclic identity, $R^m_{\phantom{m}[jkl]}=-2S^m_{\phantom{m}[jk;l]}+4S^m_{\phantom{m}n[j}S^n_{\phantom{n}kl]}$ \cite{Niko,rev,Scho}.
The curvature tensor can be decomposed as $R^i_{\phantom{i}klm}=P^i_{\phantom{i}klm}+C^i_{\phantom{i}km:l}-C^i_{\phantom{i}kl:m}+C^j_{\phantom{j}km}C^i_{\phantom{i}jl}-C^j_{\phantom{j}kl}C^i_{\phantom{i}jm}$, where $P^i_{\phantom{i}klm}$ is the Riemann tensor and colon denotes the covariant derivative with respect to the Levi-Civita connection.
The Ricci tensor is given by $R^a_{\phantom{a}i}=R^{ab}_{\phantom{ab}ij}e^j_b$ or $R_{ik}=R^j_{\phantom{j}ijk}$.

The simplest and most natural gravitational Lagrangian density is linear in the curvature tensor:
\begin{equation}
\mathfrak{L}_\textrm{g}=-\frac{1}{2\kappa}\mathfrak{e}R,
\label{ecsk}
\end{equation}
where $R=R^b_{\phantom{b}j}e^j_b=R^i_{\phantom{i}i}$ is the Ricci scalar and $\kappa=8\pi G/c^4$ is Einstein's gravitational constant (which sets the units of mass).
Such a function has no free parameters.
Varying the total action for the gravitational field and matter, $S=(1/c)\int(\mathfrak{L}_\textrm{g}+\mathfrak{L}_\textrm{m})d\Omega$, with respect to the torsion tensor (or the spin connection) and equaling this variation to zero gives the Cartan field equations \cite{Niko,KS,Hehl,rev}
\begin{equation}
S^j_{\phantom{j}ik}-S_i \delta^j_k+S_k \delta^j_i=-\frac{\kappa}{2\mathfrak{e}}\mathfrak{S}^{\phantom{ik}j}_{ik}.
\label{Cartan}
\end{equation}
These equations are linear and algebraic: torsion is proportional to the intrinsic spin density of matter and thus vanishes outside material bodies.
Varying the total action $S$ with respect to the tetrad and equaling this variation to zero gives the Einstein field equations \cite{Niko,KS,Hehl,rev}
\begin{equation}
R_{ki}-\frac{1}{2}Rg_{ik}=\frac{\kappa}{\mathfrak{e}}\mathfrak{T}_{ik}.
\label{Einstein}
\end{equation}

Substituting the field equations (\ref{Cartan}) and (\ref{Einstein}) into the contracted Bianchi identity gives the conservation law for the dynamical energy-momentum density: $\mathfrak{T}^{ij}_{\phantom{ij}:j}=C_{jk}^{\phantom{jk}i}\mathfrak{T}^{jk}+(1/2)\mathfrak{S}_{klj}R^{klji}$, which generalizes (\ref{cemd2}).
Substituting (\ref{Cartan}) and (\ref{Einstein}) into the contracted cyclic identity leads to the conservation law (\ref{amd4}) for the spin density.
A more complicated Lagrangian density for the gravitational field would give more complicated field equations.
Those equations, however, upon substituting into the contracted Bianchi and cyclic identities would still give the same conservation laws.
The conservation law for the spin density is therefore contained in the cyclic identity for the curvature tensor.
In the metric formulation of gravity, the cyclic identity reduces to $P^m_{\phantom{m}[jkl]}=0$, which leads to $\mathfrak{T}_{ij}=\mathfrak{T}_{ji}$.
This symmetry relation is consistent with (\ref{amd1}) only if the intrinsic spin is absent.
The Bianchi and cyclic identities in the metric formulation therefore contain only one independent conservation law, for the energy-momentum density, from which the conservation law for the orbital angular momentum density follows.

Varying the total action $S$ with respect to the metric tensor and equaling this variation to zero gives the Riemannian form of the Einstein equations, $G_{ik}=\kappa(T_{ik}+U_{ik})$, where $G_{ik}=P^j_{\phantom{j}ijk}-(1/2)P^{lm}_{\phantom{lm}lm}g_{ik}$ is the Einstein tensor and $U^{ik}=\kappa\bigl(-s^{ij}_{\phantom{ij}[l}s^{kl}_{\phantom{kl}j]}-(1/2)s^{ijl}s^k_{\phantom{k}jl}+(1/4)s^{jli}s_{jl}^{\phantom{jl}k}+(1/8)g^{ik}(-4s^l_{\phantom{l}j[m}s^{jm}_{\phantom{jm}l]}+s^{jlm}s_{jlm})\bigr)$ is a contribution to the energy-momentum tensor from torsion, which is quadratic in the spin tensor \cite{rev}.
The spin tensor also appears in $T_{ik}$ because $\mathfrak{L}_\textrm{m}$ depends on torsion.
The metric-affine formulation of gravity, based on the Lagrangian density (\ref{ecsk}), constitutes the Einstein-Cartan-Sciama-Kibble (ECSK) theory \cite{Niko,KS,Hehl,rev}, and the corresponding metric formulation is the standard, Einstein-Hilbert form of the general theory of relativity (GR) \cite{LL}.
Since the metric-affine formulation of gravity is required by the existence of intrinsic spin, the ECSK theory is a more complete form of GR.
The quantity $U_{ik}$ is significant only at extremely high densities, for which the square of the density of spin is on the order of the energy density multiplied by $\kappa$ \cite{non}.
In vacuum, where torsion and $U_{ik}$ vanish, both theories have the same field equations, $G_{ik}=\kappa T_{ik}$, and thus give the same predictions.
The ECSK theory of gravity therefore passes all observational and experimental tests of GR \cite{exp}.

\paragraph{Dirac spinors in spacetime with torsion.}
Elementary particles, that are fermions, are described by Dirac spinor fields (wave functions).
In the metric-affine formulation of gravity, the Dirac Lagrangian density for a free spinor $\psi$ with mass $m$, minimally coupled to the gravitational field, is given by $\mathfrak{L}_\textrm{m}=(i/2)\hbar c\mathfrak{e}(\bar{\psi}\gamma^k\psi_{;k}-\bar{\psi}_{;k}\gamma^k\psi)-mc^2\mathfrak{e}\bar{\psi}\psi$, where $\bar{\psi}=\psi^{\dag}\gamma^0$ is the adjoint spinor corresponding to $\psi$ \cite{Niko,rev}.
The covariant derivative of $\psi$, $\psi_{;k}=\psi_{,k}-\Gamma_k\psi$, gives $\bar{\psi}_{;k}=\bar{\psi}_{,k}+\bar{\psi}\Gamma_k$.
The Dirac matrices $\gamma^a$ obey $\gamma^{(a}\gamma^{b)}=\eta^{ab}I$ and transform under local Lorentz transformations like $\psi\bar{\psi}$.
The last relation yields $\gamma^a_{\phantom{a}|k}=\omega^{a}_{\phantom{a}bk}\gamma^b-[\Gamma_k,\gamma^a]$, which gives the Fock-Ivanenko spinor connection $\Gamma_k=-(1/4)\omega_{abk}\gamma^a\gamma^b$, in accordance with the generators of the Lorentz group in the spinor representation, $G^{ab}=(1/2)\gamma^{[a}\gamma^{b]}$.
Varying the total action for the gravitational field and fermionic matter with respect to the adjoint spinor $\bar{\psi}$ and equaling this variation to zero gives the Dirac equation $i\hbar\gamma^k\psi_{;k}=mc\psi$.

The energy-momentum tensor for a Dirac field is $T_{ik}=(i/2)\hbar c(\bar{\psi}\delta^j_{(i}\gamma_{k)}\psi_{;j}-\bar{\psi}_{;j}\delta^j_{(i}\gamma_{k)}\psi)-(i/2)\hbar c(\bar{\psi}\gamma^j\psi_{;j}-\bar{\psi}_{;j}\gamma^j\psi)g_{ik}+mc^2\bar{\psi}\psi g_{ik}$.
The covariant derivative of a spinor can be decomposed into the Riemannian covariant derivative and a term containing $C_{ijk}$: $\psi_{;k}=\psi_{:k}+(1/4)C_{ijk}\gamma^{[i}\gamma^{j]}\psi$, $\bar{\psi}_{;k}=\bar{\psi}_{:k}-(1/4)C_{ijk}\bar{\psi}\gamma^{[i}\gamma^{j]}$.
The contortion tensor therefore appears in the Dirac Lagrangian density in a term $(i/8)\hbar c\bar{\psi}(\gamma^k\gamma^{[i}\gamma^{j]}+\gamma^{[i}\gamma^{j]}\gamma^k)\psi C_{ijk}$.
Consequently, the spin tensor for a Dirac field is completely antisymmetric: $s^{ijk}=-(1/\mathfrak{e})\epsilon^{ijkl}s_l$, where $\epsilon^{ijkl}$ is the Levi-Civita permutation symbol, $s^i=(1/2)\hbar c\bar{\psi}\gamma^i\gamma^5\psi$ is the Dirac spin pseudovector, and $\gamma^5=i\gamma^0\gamma^1\gamma^2\gamma^3$ \cite{Niko,KS,Hehl,rev}.
This spin tensor does not depend on $m$ and remains the same if we include the electromagnetic, weak or strong interactions of fermions.
Since $C_{ijk}$ appears only in the additive, kinetic term in the Lagrangian density, $(i/2)\hbar c\mathfrak{e}(\bar{\psi}\gamma^k\psi_{;k}-\bar{\psi}_{;k}\gamma^k\psi)$, the spin density is also additive.
Accordingly, the spin tensor for a system of fermions is also completely antisymmetric.

Substituting the spin tensor for a Dirac field into the Cartan equations (\ref{Cartan}) gives the completely antisymmetric torsion tensor: $S_{ijk}=C_{ijk}=(1/2)\kappa e_{ijkl}s^l$, and the Dirac equation turns out to be nonlinear (cubic) in $\psi$: $i\hbar\gamma^k\psi_{:k}=mc\psi-(3/8)\hbar^2 c\kappa(\bar{\psi}\gamma^k\gamma^5\psi)\gamma_k\gamma^5\psi$ \cite{HD}.
The corresponding combined energy-momentum tensor is given by $T_{ik}+U_{ik}=(i/2)\hbar c(\bar{\psi}\delta^j_{(i}\gamma_{k)}\psi_{:j}-\bar{\psi}_{:j}\delta^j_{(i}\gamma_{k)}\psi)+(3/4)\kappa s^l s_l g_{ik}$.
The second term in this tensor removes the unphysical big-bang singularity, which appears in the metric GR, by a cusp-like bounce at a finite minimum scale factor, before which the Universe was contracting \cite{cosm} (a similar nonsingularity is shown in \cite{other}).
The dynamics of the Universe after the bounce also explains why the observable Universe at largest scales appears spatially flat, homogeneous and isotropic, without needing cosmic inflation \cite{cosm}.

\paragraph{Teleparallel gravity.}
The teleparallel formulation of gravity constrains the curvature tensor $R^i_{\phantom{i}mjk}$ to be zero, which is satisfied if the affine connection is equal to the Weitzenb\"{o}ck connection, $\Gamma^{\,\,j}_{i\,k}=e^a_{i,k}e^j_a$ \cite{tp}.
In this formulation, as in GR, only the tetrad is a dynamical variable in varying the action.
The corresponding spin connection (\ref{spincon}) vanishes, so another expression for the teleparallel spin connection is needed to couple spinors and the gravitational field.
Such a coupling is thus nonminimal.
The expression $\omega^a_{\phantom{a}bk}=-C^a_{\phantom{a}bk}$ is consistent with GR; it satisfies $\delta\omega^{ab}_{\phantom{ab}i}=-\epsilon^{ab}_{\phantom{ab}|i}$ for infinitesimal Lorentz transformations \cite{tpspin}.
Since the teleparallel contortion tensor is a function of the tetrad and its first derivatives, the spin density (\ref{spind2}) is thus a function of the energy-momentum density, forming a part of the orbital angular momentum density, whereas (\ref{spind1}) is no longer valid.

A teleparallel gravitational Lagrangian density, which gives the same field equations as GR, is given by $\mathfrak{L}_\textrm{g}=(1/2)(\mathfrak{e}/\kappa)T$, where $T=S^{ijk}S_{ijk}+2S^{ijk}S_{jik}-4S^i S_i$ \cite{tp}.
The teleparallel formulation of gravity, based on this Lagrangian density, constitutes the teleparallel equivalent of general relativity (TEGR).
This expression is less fundamental than the simple Lagrangian density (\ref{ecsk}) of the ECSK theory; it has two free parameters that were chosen to give the desired field equations.
The contracted cyclic identity leads to the conservation law for the dynamical energy-momentum density, $\mathfrak{T}^{ij}_{\phantom{ij}:j}=0$, whereas the contracted Bianchi identity is satisfied by construction.
The field equations impose the symmetry condition on (\ref{demd1}), $\mathfrak{T}_{ij}=\mathfrak{T}_{ji}$ \cite{tpDir}, which is not a generalization of (\ref{amd3}) unless the intrinsic spin vanishes.
The symmetry condition arises also in other teleparallel theories of gravity, such as $f(T)$ gravity \cite{ttg}.
Absolute parallelism therefore excludes the intrinsic spin.
Consequently, the observed existence of matter with intrinsic spin requires a nonzero curvature tensor, ruling out the teleparallel formulation of gravity.

\paragraph{Summary.}
The observed existence of matter with intrinsic spin requires spacetime to be equipped with both curvature and torsion.
Regarding the tetrad and spin connection as independent variables gives the correct conservation law for the total (orbital plus intrinsic) angular momentum in the presence of the gravitational field.
Extending GR into the simplest theory of gravity with curvature and torsion, the ECSK theory, not only includes the intrinsic spin but also avoids the big-bang singularity.
Torsionless theories, such as metric $f(R)$ gravity \cite{fR}, and teleparallel theories, such as TEGR \cite{tp} and $f(T)$ gravity, are structurally inconsistent with the presence of intrinsic spin and thus are unphysical.
Cosmologies based on those theories \cite{ttg} cannot be used, especially in the very early Universe where intrinsic spin is significant.

\end{document}